\documentclass[twocolumn,showpacs,preprintnumbers,amsmath,amssymb]{revtex4}

\usepackage{graphicx}
\usepackage{dcolumn}
\usepackage{bm}
\def\Vec#1{\mbox{\boldmath $#1$}}
\def\kv{\mbox{\boldmath $ k$}}

\begin{document}


\title{
Kramer-Pesch approximation for analyzing field-angle-resolved
  measurements made in unconventional superconductors: A
  calculation of the zero-energy density of states
}

\author{Yuki Nagai$^{1,2}$}
\author{Nobuhiko Hayashi$^{3,4}$}
\affiliation{
$^{1}$Department of Physics, University of Tokyo, Tokyo 113-0033, Japan\\
$^{2}$Condensed Matter Theory Laboratory, RIKEN, Saitama, 351-0198, Japan\\
$^{3}$CCSE, Japan  Atomic Energy Agency, 6-9-3 Higashi-Ueno, Tokyo 110-0015, Japan\\
$^{4}$CREST(JST), 4-1-8 Honcho, Kawaguchi, Saitama 332-0012, Japan
}

\date{\today}

\begin{abstract}
By measuring angular-oscillation behavior of the heat capacity with respect to the applied field direction, 
one can detect the details of the gap structure. 
We introduce the Kramer-Pesch approximation (KPA) as a new method to analyze
the field-angle-dependent experiments 
quantitatively.
We calculate the zero energy density of states
for various combinations of typical Fermi surfaces and superconducting gaps.
The KPA yields a merit that one can quantitatively
compare theoretical calculations with experimental results
without involving heavy numerical computations,
even for complicated Fermi surfaces.
We show an inadequacy of the frequently-used Doppler-shift technique,
which is remedied by application of the KPA.
\end{abstract}

\pacs{
74.20.Rp, 
74.25.Op, 
74.25.Bt  
}
\maketitle
Much attention has been paid to investigation on
the mechanism of
exotic superconductivity emerged in
highly-correlated materials like high-$T_{\rm c}$ cuprates and heavy-fermion compounds,
in metallic materials like nonmagnetic borocarbides,
and in organic materials like BEDT-TTF salts. 
Many exotic superconductors (SCs) have zeros of
the single-particle excitation gap called nodes on the Fermi surface (FS). 
It is then important to identify the detailed gap structure; either 
line or point nodes and their locations on the FS. 
The identification of them
is crucial for understanding
the unconventional pairing mechanism of exotic SCs~\cite{Sigrist}.

To this end, various experimental methods have been developed.
The power-law temperature dependence of
physical quantities such as specific heat and nuclear magnetic relaxation rate
provides
partial information on the gap structure, namely a difference between line and point nodes.
In the last decade,
the angular dependence on an applied magnetic field 
has facilitated us deducing more directly the gap structure
including the information on the positions of gap nodes \cite{Sakakibara,Matsuda}.
One of such powerful methods is the angular-resolved
heat capacity measurement \cite{Sakakibara}. 
By measuring
angular-oscillation behavior of
the heat capacity
$C({\bm H})$
with respect to the applied field ${\bm H}$ direction,
one
can detect the details of the gap structure, especially for the
locations of its nodes.
The angular-resolved thermal conductivity is also such an experimental technique \cite{Matsuda}. 
These measurements are capable of achiving high precision enough to discuss details of the oscillatory structure. 
Owing to potential abilities
to reveal the gap structure
in much more various exotic SCs,
experimental and theoretical progresses in this direction
are highly
influential
developments in the condensed matter physics.

For instance,
it is believed that YNi$_2$B$_2$C is
the first important material in which
superconductivity with a point-node gap
was confirmed experimentally \cite{Izawa,Park}. 
One of the evidences of point nodes in this material has been considered
to be an experimental result
that the polar-angle dependence of
the thermal conductivity is strong
as shown by Izawa {\it et al}.\ in Fig.~2 of Ref.~\cite{Izawa}
(and in Fig.~7 of Ref.~\cite{Matsuda}).
They proposed that the polar-angle dependence would be weak
if YNi$_2$B$_2$C had line nodes,
instead of point nodes,
as shown in Fig.~4(b) of Ref.~\cite{Matsuda}.
Their proposition is based on
a Doppler-shift (DS) analysis \cite{Matsuda,Izawa} for an ellipsoidal FS \cite{Izawap}.
The DS method,
which was crucial for that proposition of point nodes in YNi$_2$B$_2$C,
has been frequently utilized to analyze
data also for other various materials \cite{Matsuda}.
As noted below, however,
the DS method faces certain difficulties, making it highly questionable for quantitative analyses,
unfortunately.
It should also be noted that YNi$_2$B$_2$C has
highly-anisotropic complicated FSs \cite{Yamauchi}
in comparison to rather simple models of FSs adopted in usual studies.

There is a twofold problem in the DS method for analyzing experimental results. 
One of the problems is an overestimation of the angular-oscillation amplitude.
The DS method gives rise to an inflated oscillation amplitude 
of the order 30 \% of the maximum amplitude \cite{Vekhter},
albeit actual oscillation amplitudes are only of the order a few percents
in many experiments \cite{Sakakibara}. 
Indeed, a full numerical calculation \cite{Miranovic}
leads to an oscillation amplitude consistent with those experiments,
suggesting an inadequacy of the DS method.
The other problem is related to an issue
whether or not a cusp-like minimum appears for the magnetic field $\Vec{H} \parallel$ 
gap-nodal direction. 
Miranovi\'c {\it et al}.\ \cite{Miranovic} showed 
that a broad (non-cusp-like) minimum appears in a full numerical calculation
for a spherical FS. 
In experiments,
the cusp-like minima do not appear in most materials \cite{Sakakibara}.
Only in YNi$_2$B$_2$C the cusps were observed \cite{Izawa,Park}.
As discussed later,
for addressing this issue 
the DS method appears to be inadequate
in materials with complicated FSs because of an approximation
inherent in its formulation.
Although full numerical calculations
could resolve both of these problems in principle, they 
require significantly long computational time
and it is hard to conduct such direct calculations
in systems with realistically complicated FSs.
The Brandt-Pesch-Tewordt (BPT) method is alternatively proposed,
which is an approximation adequate near the upper critical fields $H_{c2}$ \cite{Vorontsov}. 
The BPT is also not an easy-to-calculate method in low magnetic fields $H$ and
at low temperatures $T$, unfortunately.
In this respect the DS method has been a favorably easy tool,
but it faces the above mentioned problems.
Therefore, it is now highly desirable to develop a new tractable method that provides
quantitatively correct results.

In this Letter, we first introduce the Kramer-Pesch approximation 
(KPA) as an efficiently applicable method to analyze
the angular-resolved experiments.
We will show that the KPA yields the zero energy density of states (ZEDOS)
consistent quantitatively with
results of direct numerical calculations.
The computational time required for the KPA
is almost the same as for the DS method
and is significantly less than for direct numerical calculations. 
Using the KPA, one can also obtain
the density of states (DOS)
in a system with 
realistically complicated FSs without involving heavy numerical computations.
In addition,
we will show that the DS method 
is inadequate and the KPA would be indispensable for analyzing experimental results, since 
some information on the FS is neglected in the DS method.

The DOS is the basis for analyzing physical quantities
such as the specific heat and the thermal conductivity \cite{Vorontsov}.
It is obtained from the regular Green function $g$
within the quasiclassical theory of superconductivity,
which is represented
by a parametrization with $a$ and $b$ as
$g = - (1-a b)/(1+ a b)$.
They follow the Riccati equations ($\hbar=1$) \cite{Schopohl}: 
\begin{eqnarray}
\Vec{v}_{\rm F} \cdot \Vec{\nabla} a + 2 \tilde{\omega}_n a + a \Delta^{\ast} a - \Delta &=& 0,
\label{eq:ar}\\
\Vec{v}_{\rm F} \cdot \Vec{\nabla} b - 2 \tilde{\omega}_n b - b \Delta b + \Delta^{\ast} &=& 0.
\label{eq:br}
\end{eqnarray}
Here, $\Vec{v}_{\rm F}$ is the Fermi velocity, 
and $i \tilde{\omega}_n = i \omega_n + (e/c) \Vec{v}_{\rm F} \cdot \Vec{A}$ with
the Matsubara frequency $\omega_n$ and the vector potential $\Vec{A}$.

First, we show what is the difference between the DS method and the KPA
from a fundamental viewpoint.
In the DS method, we separate the pair potential $\Delta$ into the amplitude
and the phase $\Phi(x,\kv)$
as $\Delta(x,\kv) = |\Delta| \exp(i \Phi)$,
and consider that the solution of Eq.~(\ref{eq:ar})
is written as $a = \exp (- i \Phi) \bar{a}$ \cite{Dahm}. 
Here, $x$ is the coordinate along the direction of
$\Vec{v}_{\rm F}$.
Assuming $\partial \bar{a} /\partial x= 0$, 
the equation can be exactly solved in an analytical way and
the solution is equivalent to that in the bulk region 
replacing the energy $\omega_n$ by the Doppler shifted energy $\omega_n + i \Vec{v}_{\rm F} \cdot m \Vec{v}_s$. 
The DS method is based on the fact that 
the energy of a quasiparticle (QP) shifts
by the Doppler-shift energy $\delta E = m \Vec{v}_{\rm F} \cdot \Vec{v}_{\rm s}$
due to circulating supercurrents \cite{Volovik}, 
where $\Vec{v}_{\rm s}$ means the superfluid velocity around a vortex
and $\Vec{v}_{\rm s} \perp {\bm H}$.
Consequently, the ZEDOS is obtained as
$\sim
{\rm Re}
\bigl\langle
|\delta E| / \sqrt{(\delta E)^2-|\Delta|^2}
\ \bigr\rangle
$ \cite{Vekhter,Volovik,NagaiY},
where the angular brackets mean averaging over both a FS and a unit cell of vortex lattice.
For $\Vec{H} \parallel$ gap-nodal direction,
near gap nodes the QPs
whose Fermi velocities are
parallel to $\Vec{H}$  ($\Vec{v}_{\rm F} \parallel \Vec{H}$)
lead to $\delta E = 0$.
As a result, those QPs are excluded from the contribution to the ZEDOS. 
Therefore, $C({\bm H})/T$, which is proportional to
the ZEDOS in the limit $T \rightarrow 0$,
has a minimum for $\Vec{H} \parallel$ gap-nodal direction \cite{Vekhter}. 
The DS method ($\partial \bar{a} /\partial x= 0$)
is an approximation neglecting the spatial variation of $|\bar{a}| = |a|$,
so that it breaks down near a vortex core, around which the pair potential $\Delta$, $a$, and $b$
substantially change.
The KPA avoids this neglect.
Indeed, Dahm {\it et al}.\ \cite{Dahm}
showed that the DS method misses important contributions 
from vortex-core states extending into the gap-nodal directions 
far away from a vortex core.

To take account of those contributions,
in the KPA we expand the Riccati equations up to first order in the impact parameter $y$
and the energy $\omega_n$
($i\omega_n \to E+i\eta$ and
$y$ is the coordinate along the direction perpendicular to $\Vec{v}_{\rm F}$) \cite{Nagai,Kramer}. 
It follows that only the Andreev bound states around a vortex core is taken into account by the KPA 
and this approximation is appropriate in the low-energy region
$|\omega_n| \ll |\Delta|$. 
The KPA is based on the fact that the delocalized states due to
the QPs near gap nodes
can be regarded as the 
Andreev bound states spreading to infinite distances, 
so that the Andreev bound states predominantly contribute to the DOS
in the low energy.
Indeed, Kopnin and Volovik \cite{Volovik}
confirmed that, for the DOS, the DS method and a calculation by considering only the bound states
(corresponding to the KPA here)
yielded results of the same order in the case of two-dimensional (2D) isotropic FS
under a magnetic field applied perpendicular to the 2D plane.
It should also be noted that
the KPA works well widely far away from a vortex core
since the relation $|\omega_n| \ll |\Delta|$ is marginally satisfied even near
gap nodes
in the low energy
and the KPA correctly describes vortex core states
extending into the gap-nodal directions \cite{Nagai}. 
Hence, the KPA appears to be adequate for calculating the low-energy DOS in the vortex state.

By the expansion procedure based on the KPA \cite{Nagai},
an analytic solution of the Riccati equations can be obtained around a vortex.
The DOS is calculated as $\langle -{\rm Re} g^{\rm R} \rangle$
with the retarded Green function $g^{\rm R}$ accordingly.
We find the new expression for the angular-resolved DOS
in general form,
\begin{equation}
N(E, \alpha_{\rm M}, \theta_{\rm M})
=
\frac{v_{\rm F0} \eta}
     {2 \pi^2 \xi_0}
\Biggl{\langle} 
\int 
\frac{ d S_{\rm F}  }
     { |\Vec{v}_{\rm F}|  }
\frac{
  \lambda
  \bigl[ \cosh
        (x/\xi_0)
  \bigr]^{\frac{-2 \lambda}{\pi h}} 
 }{(E-E_y)^2 + \eta^2} 
\Biggl{\rangle}_{\rm SP}.
\label{eq:dos}
\end{equation}
Here,
the azimuthal (polar) angle of ${\bm H}$ is $\alpha_{\rm M}$ ($\theta_{\rm M}$)
in a spherical coordinate frame fixed to crystal axes,
and
$d S_{\rm F}$ is an area element on the FS
[e.g.,
$d S_{\rm F}=k_{\rm F}^2 \sin\theta d\phi d\theta$
for a spherical FS
in the spherical coordinates $(k,\phi,\theta)$,
and
$d S_{\rm F}=k_{{\rm F}ab} d\phi dk_c$
for a cylindrical FS
in the cylindrical coordinates $(k_{ab},\phi,k_c)$].
In the cylindrical coordinate frame $(r,\alpha,z)$ with
${\hat z} \parallel {\bm H}$ in the real space,
the pair potential is $\Delta \equiv \Delta_0 \Lambda(\kv_{\rm F}) \tanh(r/\xi_0) \exp(i\alpha)$
around a vortex,
$\Delta_0$
is the maximum pair amplitude in the bulk, $\lambda = |\Lambda|$,
and
$\langle \cdots \rangle_{\rm SP}
\equiv \int_0^{r_a} r dr \int_0^{2 \pi} \cdots d \alpha/ (\pi r_a^2)$
is the real-space average around a vortex,
where $r_a/\xi_0 = \sqrt{H_{c2}/H}$
[$H_{c2} \equiv \Phi_0 / (\pi \xi_0^{2})$, $\Phi_0 = \pi r_a^2 H$].
$x=r\cos(\alpha -\theta_v)$,
$y=r\sin(\alpha -\theta_v)$,
and
$E_y = \Delta_0 \lambda^2 y/(\xi_0 h)$.
$\theta_v(\kv_{\rm F},\alpha_{\rm M}, \theta_{\rm M})$ 
is the angle of $\Vec{v}_{\rm F \perp}$ in the plane of $z=0$,
where $\alpha$ and $\theta_v$ are measured from a common axis \cite{Nagai,NagaiY}.
$\Vec{v}_{\rm F \perp}$
is the vector component of $\Vec{v}_{\rm F} (\kv_{\rm F})$
projected onto the plane normal to
${\hat {\bm H}}=(\alpha_{\rm M}, \theta_{\rm M})$.
$|\Vec{v}_{\rm F \perp}(\kv_{\rm F},\alpha_{\rm M}, \theta_{\rm M})|
\equiv v_{\rm F0}(\alpha_{\rm M}, \theta_{\rm M}) h(\kv_{\rm F},\alpha_{\rm M}, \theta_{\rm M})$
and 
$v_{\rm F0}$ is the FS average of $|\Vec{v}_{\rm F \perp}|$ \cite{NagaiY}.
$\xi_0$ is defined as
$\xi_0 = v_{{\rm F}0}/(\pi \Delta_0)$.
We consider here a clean SC in the type-II limit.
The impurity effect can be incorporated through the smearing factor $\eta$.
We should note that
the expression by the KPA includes
only the contribution of the Andreev states as mentioned above
and it excludes the bulk DOS under the zero magnetic field.
This is, however, sufficient for considering
the ${\bm H}$ dependence of the DOS in low energies $E$
because the bulk DOS does not depend on ${\bm H}$
and the Andreev states dominate in low $E$.
While Eq.~(\ref{eq:dos}) might seem more complicated than the DS expression,
a load of required numerical integrations is of the same order
and is much less than direct numerical calculations of the Riccati equations efficiently.

\begin{figure}
\includegraphics[width = 6cm]{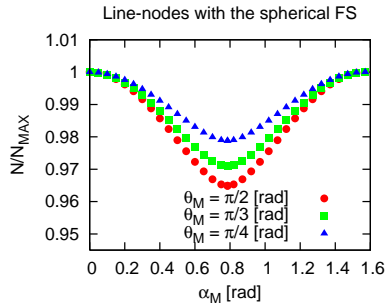}
\caption{\label{fig:fig1}
(Color online)
Angular dependence of the ZEDOS calculated by the KPA 
in a $d_{x^2-y^2}$-wave SC with the spherical FS. 
Circle $\circ$, box $\Box$, and triangle $\triangle$ indicate
the results for 
the polar angles $\theta_{\rm M} = \pi/2$, $\pi/3$, and $\pi/4$, respectively. 
The smearing factor $\eta = 0.05 \Delta_0$ and the cut-off length $r_a = 7 \xi_0$.
}
\end{figure}

\begin{figure}
  \begin{center}
    \begin{tabular}{p{40mm}p{40mm}}
      \resizebox{55mm}{!}{\includegraphics{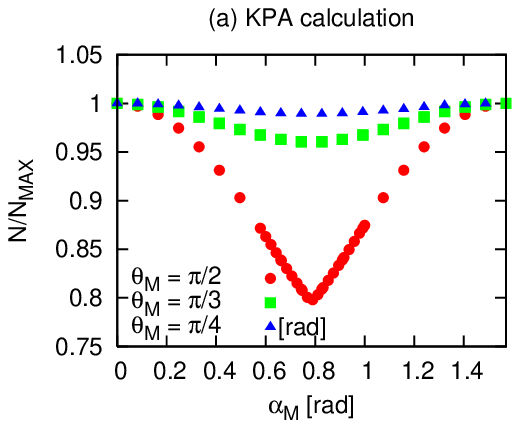}} &
      \resizebox{55mm}{!}{\includegraphics{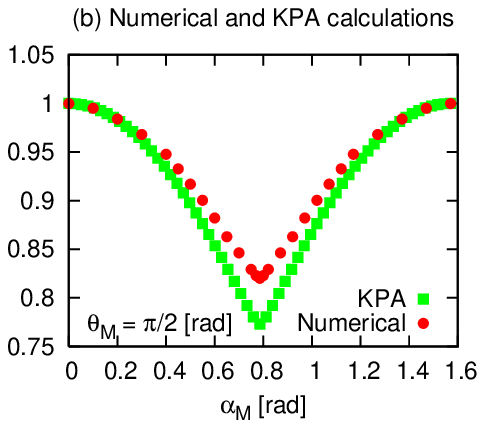}} 
    \end{tabular}
\caption{\label{fig:fig2}
(Color online)
Angular dependence of the ZEDOS obtained by (a) the KPA and
(b) the direct numerical and KPA calculations  
in the $d_{x^2-y^2}$-wave SC with the cylindrical FS. 
(a) The magnetic field tilts from the $c$ axis by polar angle $\theta_{\rm M}$.
$\eta = 0.05 \Delta_0$ and
$r_a = 7 \xi_0$. 
(b)
$\theta_{\rm M} = \pi/2$,
$\eta = 0.001 \Delta_0$, and
$r_a = 5 \xi_0$.
}
  \end{center}
\end{figure}

We show results for two cases comparing the KPA
with direct numerical calculations
to prove the quantitative correctness of the KPA.
We consider a $d_{x^2-y^2}$-wave SC. 
First, we calculate the ZEDOS ($E=0$) by the KPA for the spherical FS
(and the $d$-wave pairing $\Lambda=\cos2\phi\sin^2\theta$).
As shown in Fig.~\ref{fig:fig1}, the cusp-like minimum does not appear
and the oscillation amplitude is of the order 3.5 \%, coinciding well with
the full numerical results by Miranovi\'c {\it et al}.\ \cite{Miranovic}
where no cusp-like minimum appears and the amplitude is of the order 3.5 \%.
We have also conducted a DS calculation under the same condition
and obtained a result that
there is indeed no cusp, but the amplitude is of the order 4.9 \% with less coincidence.
Second, we calculate the ZEDOS by the KPA for the cylindrical FS
(and $\Lambda=\cos2\phi$). 
As shown in Figs.~\ref{fig:fig2}(a) and \ref{fig:fig2}(b),
the cusp-like minimum appears and the oscillation amplitude is of the order 20--23 \%.
We also calculate the ZEDOS
numerically by solving directly the Riccati equations 
to compare with the KPA result,
and find that
the cusp structure appears and the oscillation amplitude is of the order 18 \%
as shown in Fig.~\ref{fig:fig2}(b).
The DS method, in contrast, gives rise to a too large amplitude of the order 30 \% \cite{Vekhter}.
In summary
the above two cases show that the results of the KPA is quantitatively equivalent to those of
the direct numerical calculations. 
We have also confirmed that
within the KPA
the dependence of the ZEDOS on the field strength shows
$H^{0.5}$ behavior \cite{Volovik}
in gap-nodal SCs.

\begin{figure}
  \begin{center}
    \begin{tabular}{p{40mm}p{40mm}}
      \resizebox{55mm}{!}{\includegraphics{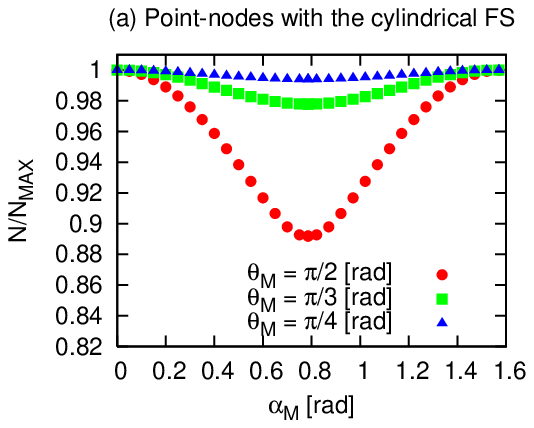}} &
      \resizebox{55mm}{!}{\includegraphics{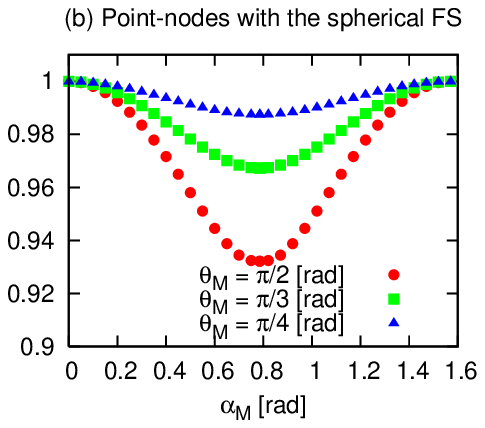}} 
    \end{tabular}
\caption{\label{fig:fig3}
(Color online)
Angular dependence of the ZEDOS calculated by the KPA 
in SCs with a point-node gap
on (a) the cylindrical and (b) the spherical FS. 
$\eta = 0.05 \Delta_0$ and
$r_a = 7 \xi_0$.
}
\end{center}
\end{figure}

We consider the polar-angle dependence of the oscillation amplitude of the ZEDOS,
motivated by the experiments on YNi$_2$B$_2$C. 
We discuss the difference of the shape of FSs and of the gap structure. 
We calculate the ZEDOS for the following four combinations;
SCs with the line-node or the point-node gap on the cylindrical or the spherical FS.
Here, the point-node gap is the $s+g$ wave \cite{Izawa}
$\Lambda= (1+\sin^4 \theta \cos 4 \phi)/2$
for the spherical FS
and is
$\Lambda= (1+\cos^4 (k_z/2) \cos 4 \phi)/2$
for the cylindrical FS.
As shown in Figs.~\ref{fig:fig1} and \ref{fig:fig3}(b),
for the spherical FS
the polar-angle dependence in the line-node case [Fig.~\ref{fig:fig1}] is different from 
that in the point-node case [Fig.~\ref{fig:fig3}(b)]. 
The polar-angle dependence 
is relatively weak in the case of the line-node gap and spherical FS [Fig.~\ref{fig:fig1}]. 
In contrast,
as shown in Figs.~\ref{fig:fig2}(a) and \ref{fig:fig3}(a),
for the cylindrical FS
the polar-angle dependence is strong 
both in the cases of line nodes and point nodes. 
These results suggest, that one cannot readily determine 
whether a SC has point nodes or line nodes
on the basis of the polar-angle dependence of angular-resolved
measurements,
since the polar-angle dependence of the ZEDOS depends more eminently on the shape of FSs
than on the type of gap nodes.

\begin{figure}
\includegraphics[width = 7cm]{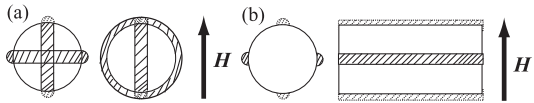}
\caption{\label{fig:fig4}Schematic figures of (a) the spherical and (b) the cylindrical FS
in the $d$-wave case. 
The left and right panel denote a top and a side view, respectively.
The hatched regions indicate the low-energy QPs around gap nodes.
}
\end{figure}

We understand the difference in the results between the spherical and cylindrical FSs 
in the $d$-wave case as follows.
The QPs near gap nodes (the nodal QPs) predominantly contribute to
the low-energy DOS.
The QPs with ${\bm v}_{\rm F}$ parallel to ${\bm H}$ are excluded from the contribution to the DOS
for the same reason as mentioned before,
so that
the ZEDOS
becomes small
when ${\bm v}_{\rm F}$ of the nodal QPs becomes parallel to ${\bm H}$.
Now, on the spherical FS, the nodal QPs with ${\bm v}_{\rm F}$  parallel to ${\bm H}$ 
are located only in narrow pole regions,
and therefore the decrease of the ZEDOS due to such a QP exclusion
is not so significant.
In contrast, 
on the cylindrical FS (see, Fig.~\ref{fig:fig4}(b))
the nodal QPs with ${\bm v}_{\rm F} \parallel {\bm H}$ are located
around long line-like regions.
The QP exclusion occurs in such wide regions,
so that
the ZEDOS dramatically decreases
for ${\bm H} \parallel$ gap-nodal direction.
As a result,
the ZEDOS prominently changes near this gap-nodal direction
and such behavior leads to the cusp-like structure
as in Fig.~\ref{fig:fig2}.
One can understand the difference in the polar-angle dependence
in the same way.

We show that the DS method is inadequate to analyze the angular-resolved experiments. 
In the DS method,
only the nodal QPs that satisfy the relation
$|m\Vec{v}_{\rm F} \cdot \Vec{v}_{\rm s}| > |\Delta|$ contribute to the ZEDOS
as noticed from the DS expression.
That is, the contribution of the anti-nodal QPs
in the region
where $|m\Vec{v}_{\rm F} \cdot \Vec{v}_{\rm s}| < |\Delta|$ is completely neglected. 
This means that the DS method overestimates
the relative contribution of the QPs near gap nodes on the FS.
As a result, the contrast in the angular-resolved ZEDOS between the nodal and anti-nodal directions is
exaggerated, leading to too large oscillation amplitudes
as mentioned before.
This problem in the DS method
would be fatal when analyzing the angular-resolved ZEDOS
for general SCs with complicated FSs
where the contributions of various FS regions
are equally expected to be important.
For example, if a FS has a narrow region where its shape is similar to the cylindrical FS,
a point-node gap in this region gives rise to unexpected cusp structure
similar to the cusped ZEDOS for a line-node gap
because the DS method considers only regions near gap nodes
and does not clearly distinguish whether those gap nodes are distributed in side-by-side array (i.e., a line node) or in isolation (i.e., point nodes).
Indeed, we have confirmed that a DS calculation under the same condition
as in Fig.\ \ref{fig:fig3}(a) gives rise to inappropriate cusp-like structure.
In contrast,
the whole region on a FS is considered by the KPA
because,
through the vortex-core states,
the KPA
takes account also of
the QPs in the anti-nodal region on the FS, which are neglected in the DS method.
In other words, the DOS obtained by the KPA includes correctly
both the contributions of the ``gap node" in the real space (i.e., the vortex core)
and the gap nodes on a FS in the momentum space.

The cusp-like minimum was observed in YNi$_2$B$_2$C \cite{Park,Izawa}. 
From the results for the typical FSs
shown in Figs.~\ref{fig:fig1}--\ref{fig:fig3},
it is anticipated that,
as realized in the case of the cylindrical FS with the line-node gap,
the cusp-like minimum appears
in the case when there exist many nodal QPs
with ${\bm v}_{\rm F}$ directed in a common direction. 
This suggests that,
if YNi$_2$B$_2$C has a point-node gap,
the directional distribution of ${\bm v}_{\rm F}$
around gap nodes should have singular structure
under favor of the highly anisotropic FSs \cite{Yamauchi} in this material. 
This anticipation would be consistent with a claim \cite{NagaiY}
that the point nodes would be situated in the regions on the FS connected by
nesting vectors. 
The importance of the FS structure has been emphasized by 
Udagawa {\it et al}.\ with the BPT calculation\cite{Udagawa}.
Further detailed analyses based on the realistic FSs are necessary to gain 
conclusive insight, for which the main result of the present work, Eq.~(\ref{eq:dos}),
applicable to general shapes of FSs
will play an efficient role.

In conclusion, we introduced the KPA as a new method that allows us
to analyze angular-resolved experiments
without heavy numerics.
According to the results for the typical FSs,
the FS shape eminently affects
the angular-resolved ZEDOS,
suggesting that
one cannot readily identify
a superconducting gap in a material
without considering its realistic FS shape.
The KPA is a useful method as a more proper approximation
in a low-$H$ and low-$T$ region than the DS technique
and is complementary to the BPT method \cite{Vorontsov} appropriate near $H_{\rm c2}$. 
We expect that the present work will stimulate further experimental efforts to detect precise oscillatory structure.

We thank Y.\ Kato, K.\ Izawa, M.\ Udagawa, P.\ Miranovi\'c, N.\ Nakai, M.\ Ichioka,
K.\ Machida, and M. Machida for helpful discussions and comments.


\begin{thebibliography}{99}
\bibitem{Sigrist}M. Sigrist and K. Ueda, Rev, Mod. Phys. {\bf 63}, 239 (1991).
\bibitem{Sakakibara}T. Sakakibara {\it et al.}, J. Phys. Soc. Jpn. {\bf 76}, 051004 (2007). 
\bibitem{Matsuda}Y. Matsuda {\it et al.}, J. Phys.: Condens. Matter {\bf 18}, R705 (2006).
\bibitem{Izawa}K. Izawa {\it et al.}, Phys. Rev. Lett. {\bf 89}, 137006 (2002).
\bibitem{Park}T. Park {\it et al.}, Phys. Rev. Lett. {\bf 90}, 177001 (2003).
\bibitem{Izawap}K. Izawa, private communication (2007).
\bibitem{Yamauchi}K. Yamauchi 
{\it et al.}, Physica C {\bf 412-414}, 225 (2004).
\bibitem{Vekhter}I. Vekhter {\it et al.}, Phys. Rev. B {\bf 59}, R9023 (1999).
\bibitem{Miranovic}P. Miranovi\'c {\it et al.}, Phys. Rev. B {\bf 68}, 052501 (2003).
\bibitem{Vorontsov}A. B. Vorontsov and I. Vekhter, Phys. Rev. Lett. {\bf 96}, 237001 (2006);
Phys. Rev. B {\bf 75}, 224501 (2007); Phys. Rev. B {\bf 75}, 224502 (2007).
\bibitem{Schopohl}N. Schopohl and K. Maki, Phys. Rev. B {\bf 52}, 490 (1995);
N. Schopohl, arXiv:cond-mat/9804064.
\bibitem{Dahm}T. Dahm {\it et al.}, Phys. Rev. B {\bf 66}, 144515 (2002).
\bibitem{Volovik}G. E. Volovik, JETP Lett. {\bf 58}, 469 (1993);
N. B. Kopnin and G. E. Volovik, JETP Lett. {\bf 64}, 641 (1996).
\bibitem{NagaiY}Y. Nagai {\it et al.}, Phys. Rev. B {\bf 76}, 214514 (2007).
\bibitem{Kramer}L. Kramer and W. Pesch, Z. Phys. {\bf 269}, 59 (1974).
\bibitem{Nagai}Y. Nagai 
{\it et al.}, J. Phys. Soc. Jpn. {\bf 75}, 104701 (2006).
\bibitem{Udagawa}M. Udagawa {\it et al.}, Phys. Rev. B {\bf 71}, 024511 (2005).
\end{thebibliography}
\end{document}